\documentclass[11pt,aps,prd,showpacs,nofootinbib,superscriptaddress,preprint,preprintnumbers]{revtex4}

\pdfoutput=1

\usepackage[usenames,dvipsnames]{color}
\usepackage{graphicx}
\usepackage{mathrsfs}
\usepackage{amsmath,amsfonts,amssymb,dsfont,bm,dcolumn,float}
\usepackage{pstricks,feyn,verbatim}
\usepackage{url,hyperref}
\usepackage{ifthen,bbm}

\usepackage{slashed,xcolor}
\usepackage{multirow,array,rotating}
\usepackage{fancybox,epstopdf}

\newcommand{\GeV}      {~\mathrm{GeV}}
\newcommand{\TeV}      {~\mathrm{TeV}}

\newcommand{\pb}      {~\mathrm{pb}}
\newcommand{\fb}      {~\mathrm{fb}}
\newcommand{\ab}      {~\mathrm{ab}}

\newcommand{\ba}{\begin{array}}
\newcommand{\ea}{\end{array}}
\newcommand{\beqn}{\begin{eqnarray}}
\newcommand{\eeqn}{\end{eqnarray}}
\newcommand{\beqs}{\begin{subequations}}
\newcommand{\eeqs}{\end{subequations}}
\newcommand{\be}{\begin{equation}}
\newcommand{\ee}{\end{equation}}
\newcommand{\non}{\nonumber \\}

\newcommand{\mathsym}[1]{{}}

\def\kpc{\rm kpc}

\def\cm{\rm cm}
\def\sr{\rm sr}
\def\sec{\rm sec}

\def\gU{\rm U}
\def\gSU{\rm SU}

\def\mL{\mathcal{L}}
\def\mM{\mathcal{M}}

\def\mO{\mathcal{O}}

\newcommand{\MET}{E\hspace{-0.1in}\slash_T}

\def\hf{\frac{1}{2}}

\begin{document}

\preprint{\font\fortssbx=cmssbx10 scaled \magstep2
\hbox to \hsize{
\hfill$\vcenter{\hbox{ MITP/15-004}
             }$}
}

\title{The leptophilic dark matter with $Z'$ interaction: \\ from indirect searches to future $e^+ e^-$ collider searches}

\author{Ning Chen}
\email{chenning@ustc.edu.cn}
\affiliation{Department of Modern Physics, University of Science and Technology of China, Hefei, Anhui, 230026, China}
\author{Jian Wang}
\email{jian.wang@uni-mainz.de}
\affiliation{PRISMA Cluster of Excellence \& Mainz Institute for Theoretical Physics,
Johannes Gutenberg University, D-55099 Mainz, Germany}
\author{Xiao-Ping Wang}
\email{xiaowang@uni-mainz.de}
\affiliation{PRISMA Cluster of Excellence \& Mainz Institute for Theoretical Physics,
Johannes Gutenberg University, D-55099 Mainz, Germany}

\begin{abstract}
We investigate the scenario where the dark matter only interacts with the charged leptons in the standard model via a neutral vector mediator $Z'$. Such a scenario with a 430 GeV dark matter can fit the recent positron fluxes observed by the AMS-02 Collaborations, with the reasonable boost factors. We study the possibility of searching such leptophilic $Z'$ via its lepton final states and invisible decay modes at the future electron-positron colliders, such as the International Linear Collider (ILC) and the Compact Linear Collider (CLIC). We find that for the benchmark models with $Z'$ mass from $1.0\,\TeV$ to $1.5\,\TeV$, the searches for the invisible decays of $Z'\to \bar \chi \chi$ is easily achieved at the CLIC $1.5\,\TeV$ runs via the mono-photon process. However, lighter $Z'$ with mass from $0.5\,\TeV$ to $0.8\,\TeV$ are challenging to see. The di-lepton plus single photon channel can reveal the $Z'$ mass at the ILC and CLIC with moderate luminosities.
\end{abstract}


\maketitle

\newpage


\section{\hspace*{-2mm}Introduction}
\label{section:introduction}

The existence of dark matter (DM) has been confirmed by many cosmological and astrophysical observations, while little is known about the nature of the DM candidates and their interactions with the Standard Model (SM) particles in the visible sector. Experimental searches to the DM candidates and their interactions with the SM particles include the underground direct searches, indirect searches and the collider searches. Among them, the most intriguing results are the recent observations of the excessive high-energy positrons that cannot be explained by the known astrophysical sources. One possible explanation for these excessive high-energy positrons is due to the DM contributions through annihilation or decay processes. The excess of the high-energy positron fluxes was first observed by PAMELA~\cite{Adriani:2008zr,Adriani:2010ib,Adriani:2013uda}. The most recent measurement of the positron energy spectrum extends the positron kinematic energy up to $300\,\GeV$~\cite{Adriani:2013uda}. Later, the observation of the excessive high-energy positrons was reported by the Alpha Magnetic Spectrometer (AMS) collaboration~\cite{Aguilar:2013qda}. More recently, the AMS-02 extended the measurements of the positron fraction up to $\sim 500\,\GeV$ with improved precision~\cite{Accardo:2014lma, Aguilar:2014mma}. It is also shown that the positron fraction no longer exhibits an increase with energy above $200\,\GeV$~\cite{Pohl:2014jea}.

There have been a lot of efforts to address the anomalous positron fluxes observed by the PAMELA and AMS-02 results, including the Sommerfeld enhancement effects~\cite{Cholis:2008hb,Cirelli:2008pk, ArkaniHamed:2008qn, Cholis:2008qq,Cholis:2008wq,Pospelov:2008jd, Feng:2009hw, Meade:2009iu, Finkbeiner:2010sm}, the lightest neutralino as DM candidate in the supersymmetric models~\cite{Bergstrom:2008gr, Grajek:2008pg, Chen:2010yi, Ibe:2013jya}, and other possibilities~\cite{Barger:2008su, Hooper:2008kg, Feldman:2008xs, Yin:2008bs,Ibarra:2008jk, Chen:2008qs, Zhang:2008tb, Profumo:2009uf,Kohri:2009yn,Cholis:2013psa,Yin:2013vaa, Ibarra:2013zia, Geng:2013nda}. Here we consider a leptophilic DM setup with a $Z'$ vector boson being the mediator between the visible sector and the dark sector. In this setup, the DM candidate annihilates into three generations of charged leptons equally through the $Z'$ interactions. We do not consider the DM annihilation channels to quarks or gluons since there are no excessive anti-proton observed in the PAMELA and AMS-02 yet~\cite{Adriani:2008zq, Adriani:2010rc}. The current direct search experiments and hadron collider searches also provide constraints to the DM effective couplings with quarks and gluons~\cite{Goodman:2010yf, Bai:2010hh, Goodman:2010ku, Rajaraman:2011wf, Fox:2011pm}. With a boost factor of ${\rm BF}\sim \mO(100)$, the DM candidate with mass of $430\,\GeV$ is likely to fit the AMS-02 observation of the excessive positron fluxes at the high energy region. This leptophilic $Z'$ together with its coupling to the DM candidate can be tested at the future high-energy $e^+ e^-$ colliders, such as the international linear collider (ILC)~\cite{Baer:2013cma} and the compact linear collider (CLIC)~\cite{Assmann:2000hg}. The direct discovery of such leptophilic $Z'$ can be most easily achieved through the opposite-sign-same-flavor (OSSF) di-lepton final state associated with a photon. We analyze the potential to discover the interaction between $Z'$ and DM in this setup at the future CLIC runs at $1.5\,\TeV$. Our results show that for $M_{Z'}\in (1.0\,,1.5)\,\TeV$, the mono-photon channel appears quite promising to detect the interaction after imposing a set of appropriate kinematic cuts.

This paper is organized as follows. In section~\ref{section:model}, we describe the setup of the simplified leptophilic DM model, where a vector boson $Z'$ mediates the DM interactions with the charged leptons via the $s-$channel. A set of benchmark models are listed where the $Z'$ vector boson masses are chosen to be $M_{Z'}\in (0.5\,\TeV\,,1.5\,\TeV)$. We also discuss the relic densities for the leptophilic DM. In section~\ref{section:indirect}, we discuss the indirect detections of DM candidates via the positron fluxes in the leptophilic $Z'$ model, and compare with the latest AMS-02 results. Section~\ref{section:collider_search} is devoted to analyze the discovery potential of the leptophilic $Z'$ at the future $e^+ e^-$ colliders such as ILC and CLIC. The discovery of the leptophilic $Z'$ is easily achieved via the OSSF di-lepton plus single photon final state, with the integrated luminosities of $\int\mL dt \sim \mO(1)\,\pb^{-1}-\mO(10)\,\fb^{-1}$ for the benchmark models. Furthermore, we analyze the searches for the DM at the CLIC $1.5\,\TeV$ runs via the mono-photon channel. Finally, we conclude in section~\ref{section:conclusion}.


\section{\hspace*{-2mm}The Model Setup}
\label{section:model}

\subsection{The leptophilic $Z'$ with the DM couplings}

For models with extended gauge symmetries such as $\gU(1)$ and/or $\gSU(2)$, a neutral vector boson $Z'$ is quite usual as a new physical state. Assuming a DM particle in the spectrum~\cite{Langacker:2000ju,Langacker:2008yv, Chiang:2011cv}, it is natural to consider the $Z'$ as the $s-$channel mediator between the visible and the dark sector. Such a $Z'$ is likely to couple with different DM candidates with various spin natures. The details of the $Z'$ couplings with different DM candidates and their annihilation cross sections can be found in Ref.~\cite{Berlin:2014tja}. For our illustration in this work, we assume a Dirac fermion $\chi$ as the DM candidate with the general couplings with the $Z'$ as follows,
\beqn\label{eq:DiracDMZp_coup}
\mL_{\rm DM}&=& \bar \chi \gamma^\mu ( g_\chi^V + g_\chi^A \gamma_5)   \chi   Z'_\mu\,.
\eeqn
Several previous studies have carried analysis of the direct, indirect, and the collider searches for the $Z'$ interactions with the DM candidates~\cite{Feldman:2007wj, Cassel:2009pu, Dudas:2009uq, Buckley:2011mm, Frandsen:2011cg, Alves:2013tqa, Arcadi:2013qia, Lebedev:2014bba, Bell:2014tta,Lee:2014tba,Alves:2015pea}. In Refs.~\cite{Alves:2013tqa, Arcadi:2013qia, Lebedev:2014bba,Alves:2015pea}, the mediator $Z'$ is assumed to couple with both SM quarks and leptons. Consequently, the constraints from the direct searches for DM should apply. There also exist strong constraints from the LHC searches for the mediator via the dijet process~\cite{Aad:2014aqa} and the DM candidates via the mono-jet process~\cite{Khachatryan:2014rra} respectively. To avoid such constraints from mono-jet process and to address the possible positron excesses from the latest AMS-02 results, we assume a leptophilic $Z'$ coupling of the following form
\beqn\label{eq:lepZp_coup}
\mL_{\ell}&=& \sum_i \bar \ell_i \gamma^\mu ( g_\ell^V + g_\ell^A \gamma_5)  \ell_i Z'_\mu\,,~~~i = 1\,,2\,,3\,,
\eeqn
with $\ell_i$ representing three generations of charged leptons. In the sequential SM case for the $Z'$ vector boson, the vectorial and axial couplings in Eq.~\eqref{eq:lepZp_coup} are given by
\beqn
&& g_\ell^V = -\frac{g}{4c_W} (1 - 4  s_W^2 )\,,\qquad g_\ell^A = \frac{g}{4c_W}  \,,
\eeqn
with the short notation of the Weinberg angle $(s_W\,, c_W)\equiv (\sin\theta_W\,, \cos\theta_W)$. In our discussions below, we simplify the $Z'$ couplings with the DM and the charged leptons as follows
\beqn
&&\lambda_\chi= g_\chi^V = g_\chi^A\,,~~~\lambda_\ell= g_\ell^V = g_\ell^A\,,
\eeqn
in Eqs.~\eqref{eq:DiracDMZp_coup} and \eqref{eq:lepZp_coup}. The $\lambda_\chi$ and $\lambda_\ell$ characterize the overall $Z'$ coupling strengths with the DM candidate $\chi$ and the charged leptons respectively.

\subsection{The experimental constraints}

There are several experimental constraints to consider for the leptophilic $Z'$ couplings given in Eq.~(\ref{eq:lepZp_coup}). The leptophilic $Z'$ couplings will contribute to the lepton anomalous magnetic dipole~\cite{Fayet:2007ua} as
\beqn\label{eq:g-2_Zp}
\Delta(g-2)_\ell&\sim&\frac{\lambda_\ell^2}{6\pi^2} \frac{m_\ell^2}{M_{Z'}^2}\,.
\eeqn
Specifically, the measurement of the muon anomalous magnetic moment from the E821 experiment at BNL~\cite{Bennett:2006fi} shows a discrepancy~\cite{Agashe:2014kda} with the SM predictions of
\beqn\label{eq:g-2_exp}
\Delta a_\mu&\equiv& a_\mu^{\rm exp} - a_\mu^{\rm SM}=288(63)(49)\times 10^{-11}\,,
\eeqn
where $a_\mu \equiv (g_\mu -2)/2$. Accordingly, the upper bound of the coupling $\lambda_\ell$ from the $g-2$ constraint is
\beqn
\lambda_\ell&\lesssim&5.6\times 10^{-3}\, \Big( \frac{M_{Z'}}{\GeV} \Big) \,.
\eeqn
This constraint is moderate for the very heavy $Z'$ with mass of $M_{Z'}\sim \mO(1)\,\TeV$.

The coupling of $\lambda_\ell$ is mostly constrained by the LEP-II searches for the $Z'$. For a vectorial coupling to electrons, this bound reads~\cite{LEP:2003aa}
\beqs\label{eq:lepZp_coup_LEPII}
\beqn
\lambda_e&\lesssim& 0.25\times \Big( \frac{M_{Z'}}{\TeV} \Big)\,,\\
\sqrt{ \lambda_\ell  \lambda_e} &\lesssim& 0.14\times \Big( \frac{M_{Z'}}{\TeV} \Big) ~~ {\rm for}~\ell\neq e\,.
\eeqn
\eeqs
The $Z'$ will be considered in the mass range of $M_{Z'}\in (0.5\,\TeV\,, 1.5\,\TeV)$ below. According to the LEP-II constraint listed in Eqs.~\eqref{eq:lepZp_coup_LEPII}, we fix the $Z'$ coupling with charged leptons as:
\beqn
\lambda_\ell =
\left\{
  \begin{array}{ll}
    0.05, & \hbox{ $M_{Z'}< 1.0\,\TeV\,$,} \\
    0.1, & \hbox{  $M_{Z'}\geq 1.0\,\TeV\,$,}
  \end{array}
\right.
\eeqn
and vary the $Z'$ coupling with the DM $\lambda_\chi$ only.

\subsection{The DM relic density}

The relic densities of the DM candidates are obtained by solving the Boltzmann equation for the evolution of the DM number density $n$
\beqn\label{eq:Boltzmann}
\frac{dn}{dt}&=& -3H n - \hf \langle \sigma v \rangle ( n^2 - n_{\rm eq}^2 )\,,
\eeqn
where $n_{\rm eq}$ denotes the DM equilibrium number density, $H$ is the Hubble parameter, $v$ is the ``relative velocity'', and $\sigma$ is the usual spin-averaged cross section~\footnote{The factor of $\hf$ in Eq.~\eqref{eq:Boltzmann} is due to the Dirac fermion as DM.}. It is well-known that the thermally averaged annihilation cross section $\langle \sigma v\rangle$ plays the key role in determining the DM relic densities. The solution of Eq.~\eqref{eq:Boltzmann} yields the DM relic density as
\beqn\label{eq:RelicDensity}
&&\Omega h^2 \simeq \frac{2.14 \times 10^9 \GeV^{-1} }{ J g_*^{1/2} M_{\rm pl}}\,,
\eeqn
where $h$ is today's Hubble parameter, $M_{\rm pl}= 1.22\times 10^{19}\,\GeV$, and $g_*$ is the number of effective degrees of freedom at the freeze-out temperature. The quantity $J$ is given by
\beqn
J&=&\int_{x_f}^\infty \frac{ \langle \sigma v \rangle }{x^2} dx\,,
\eeqn
where $x\equiv m/T$ and $x_f$ is the value given at the freeze-out temperature. In the non-relativistic limit, the thermally averaged annihilation cross section is given by~\cite{Gondolo:1990dk}
\beqn\label{eq:thermal_sigmaV}
\langle \sigma v\rangle &\simeq&\frac{2x^{3/2}}{\sqrt{\pi} } \int_0^\infty (\sigma  v) \epsilon^{1/2} \exp(-x\epsilon) d\epsilon\,,
\eeqn
with $\epsilon= (s-4m_\chi^2)/(4m_\chi^2)$ and
\beqn
\sigma v&=& \frac{1}{64\pi^2 (s-2 m_\chi^2)} \sqrt{1-\frac{4m_f^2}{s}} \int d\Omega\,|\overline \mM|^2\,,
\eeqn
where $|\overline \mM|^2$ is the spin-averaged amplitude square. Far from the resonance and particle production threshold, one can often expand the annihilation cross sections in powers of $\epsilon$,
\beqn\label{eq:annxsec_expansion}
\sigma v&\approx& a + b \epsilon+\mO(\epsilon^2)\,.
\eeqn
Making use of Eq.~\eqref{eq:thermal_sigmaV}, one obtains the thermally averaged annihilation cross section
\beqn
\langle \sigma v \rangle&\simeq& a + \frac{3}{2} b\,x^{-1} + \mO(x^{-2})\,.
\eeqn
Thus, when the annihilation occurs not close to a resonance, the relic density can be approximated by
\beqn\label{eq:RD_approx}
\Omega h^2 &\simeq& \frac{ 2.14\times 10^9 \GeV^{-1}\, x_f }{g_*^{1/2} M_{\rm pl} (a+3b/4 x_f) }\,.
\eeqn
However, when the annihilation is near the resonance, one should determine $J(x_f)$ by performing the numerical integral of
\beqn
J(x_f)&=&\int_0^\infty dv\, \frac{v^2 (\sigma  v)}{\sqrt{4\pi}} \int_{x_f}^\infty dx\, x^{-1/2} \exp(-xv^2/4)\,.
\eeqn
The details of the near resonance DM annihilations are given in Refs.~\cite{Griest:1990kh,Gondolo:1990dk}.

In our setup, there are two primary DM annihilation channels depending on the mass relations between $M_{Z'}$ and $m_\chi$:

\begin{enumerate}

\item For $m_\chi < M_{Z'}$, the DM annihilate into the SM lepton pairs through the $s-$channel $Z'$ exchanges. 

\item For $m_\chi > M_{Z'}$, the DM either annihilate into the SM lepton pairs through the $s-$channel $Z'$ exchanges; or annihilate into the $Z'$ pairs through the $t-$channels.

\end{enumerate}

\begin{table}[htb]
\begin{center}
\begin{tabular}{c|c|c|c|c|c}
\hline
$M_{Z'}$ [$\TeV$]  & $\lambda_\chi$ & $\Omega_{\rm CDM}h^2$ & BF & $\Gamma_{\rm vis} (\GeV)$ &  $\Gamma_{\rm inv} (\GeV)$ \\\hline\hline
$0.5$ & $1.1$ & $0.12$ & $154$  & $0.2$ & 0 \\\hline
$0.6$ & $0.8$ & $0.12$ & $175$ & $0.2$ & 0 \\\hline
$0.7$ & $0.6$ & $0.11$ & $134$  & $0.3$ & 0 \\\hline
$0.8$ & $0.24$ & $0.12$ & $134$  & $0.3$ & 0 \\\hline
\hline
 $1.0$ & $0.20$ & $0.13$ & $330$  & $1.6$ & $0.8$ \\\hline
 $1.1$ & $0.44$ & $0.12$ & $223$  & $1.8$ & $6.0$ \\\hline
 $1.2$ & $0.68$ & $0.12$ & $207$  & $1.9$ & $17.7$ \\\hline
 $1.3$ & $0.90$ & $0.13$ & $217$  & $2.1$ & $37.3$ \\\hline
 $1.4$ & $1.23$ & $0.12$ & $191$   & $2.2$ & $79.7$ \\\hline
 $1.5$ & $1.52$ & $0.12$ & $192$  & $2.4$ & $138.4$ \\\hline
 \hline
\end{tabular}
\caption{\label{tab:DM_benchmark} The benchmark models with Dirac DM mass of $m_\chi = 430 \,\GeV$ and different $M_{Z'}$ inputs. The couplings of $\lambda_\chi$ are chosen so that the corresponding relic densities are $\Omega_{\rm CDM}h^2\simeq 0.12$. The BF denotes the boost factor needed to fit the positron fluxes of the AMS-02 experiments. The last two columns are the decay widths of $Z'$, calculated by Eqs.~(\ref{eq:Zpvis_wid}) and (\ref{eq:Zpinvis_wid}). }
\end{center}
\end{table}

In our analysis, we only consider the cases where the DM annihilations occur not close to the resonance. Hence, the approximate results of Eq.~\eqref{eq:RD_approx} is sufficient for evaluating the DM relic densities. To get the benchmark models, one should consider the DM relic densities which are related to the DM annihilation cross sections. Making use of the approximate expression of the relic density in Eq.~(\ref{eq:RD_approx}), we have
\beqn\label{eq:RD_lepZpDM}
\Omega h^2&\simeq& \frac{ 1.07 \times 10^9 \GeV^{-1} }{\sqrt{g_*} M_{\rm pl} } \frac{x_f}{1 + \delta/x_f}\frac{\pi (M_{Z'}^2 - 4 m_\chi^2 )^2}{3  \lambda_\chi^2 \lambda_\ell^2 m_\chi^2 }\,,
\eeqn
where a phase space factor $\delta$ reads
\beqn
\delta&=& \frac{1 + 20\,m_\chi^2 / M_{Z'}^2 }{4(1 - 4\, m_\chi^2 / M_{Z'}^2 )}\,.
\eeqn

The current measurements of the cold DM relic density from both WMAP~\cite{Hinshaw:2012aka} and Planck satellite~\cite{Ade:2013ktc} give $\Omega_{\rm CDM}h^2  \simeq 0.12$, which will be considered as the major constraint for us to obtain the benchmark models. In Table.~\ref{tab:DM_benchmark}, we list the leptophilic DM benchmark models for the further studies. Here, we choose the DM mass as $m_\chi= 430\,\GeV$ in order to fit the current AMS-02 positron flux spectrum, to be discussed in the next section. The masses of $Z'$ vector boson are taken in the range of $M_{Z'}\in (0.5\,\TeV\,, 1.5\,\TeV)$, where we avoid the near-resonance choice of $M_{Z'}=0.9\,\TeV$. The coupling strength $\lambda_\chi$ between $Z'$ and DM for each case is taken as in Table.~\ref{tab:DM_benchmark} so that the corresponding DM relic density fits the observed value of $\Omega h^2\simeq 0.12$. The evaluation of the DM relic densities here are made by {\tt micrOMEGAs}~\cite{Belanger:2013oya}. The details of other elements in the table will be given in the following context of the indirect detections and the collider searches.


\section{\hspace*{-2mm}The Indirect Detections of DM}
\label{section:indirect}

In this section, we discuss the indirect searches for the DM candidates following the model setup above. In our case, the DM candidate annihilates into three generations of charged lepton pairs $\ell_i^+ \ell_i^-$ equally. Besides of the $e^+ e^-$ final states, the $(\mu^+ \mu^- \,, \tau^+ \tau^-)$ final states also contribute to the positron energy spectrum due to their decays. Some details can be found in Refs.~\cite{Ciafaloni:2010ti, Cirelli:2010xx}. The more recent work on fitting the AMS-02 results into various leptonic final states include Refs.~\cite{DeSimone:2013fia, Lin:2014vja, Cao:2014cda, Jin:2014ica}, where the DM annihilations into different leptonic final states were considered separately. The positron propagation in the Milky Way is described by the following diffusion equation~\cite{Ibarra:2008qg}:
\beqn\label{eq:positron_diffusion}
&&\nabla \cdot [ K(E\,,\vec r) \nabla f_+] + \frac{\partial }{\partial E} [ b(E\,,\vec r) f_+ ] + Q(E\,,\vec r) = 0 \,,
\eeqn
where $f_+$ is the number density of positron per unit energy, $K(E\,,\vec r)$ is the diffusion coefficient, $b(E\,,\vec r)$ is the rate of the energy loss, and $Q(E\,,\vec r)$ describes the source of positron. The primary positron sources are due to the DM annihilation
\beqn\label{eq:positron_source}
&& Q(E\,,\vec r) = \frac{1}{2} \Big( \frac{\rho(\vec r)}{m_\chi} \Big)^2 \sum_i \langle \sigma v \rangle_i \Big( \frac{dN_+ }{dE} \Big)_i\,,
\eeqn 
where the summation goes through all possible channels that generate positrons in the final states. The primary positron flux from the DM annihilations is given by
\beqn\label{eq:prime_positron_flux}
\Phi_+^{\rm prim}(E)&=&\frac{c}{4\pi} f_+(E\,, r_\odot)\,,
\eeqn
where $r_\odot \sim 8.5 \,\kpc$ is the distance between the Sun and the galactic center. In practice, we obtain the {\tt CalcHEP}~\cite{Belyaev:2012qa} model files for the benchmark model by using the {\tt LanHEP} package~\cite{Semenov:1996es}. These model files will be passed to {\tt micrOMEGAs}~\cite{Belanger:2013oya} in order to evaluate the primary positron fluxes. The DM density $\rho(r)$ is chosen to be the local density at the Sun $\rho_\odot\approx 0.3\,\GeV/\cm^3$ and the halo profile function $F(r)$:
\beqn\label{eq:DM_density}
\rho(r)&=&\rho_\odot F(r)\,,
\eeqn
where the default Zhao profile~\cite{Zhao:1995cp} in the {\tt micrOMEGAs} is taken. In addition, there are secondary positrons due to interactions between cosmic rays and interstellar medium, whose flux can be well approximated as~\cite{Moskalenko:1997gh,Baltz:1998xv}:
\beqn\label{eq:second_positron_flux}
\Phi_+^{\rm sec}(E)&=& \frac{4.5\, E^{0.7} }{1+650\,E^{2.3} + 1500\, E^{4.2} } ~~~[\GeV\cdot \cm^2\cdot \sec \cdot \sr ]^{-1}\,.
\eeqn

\begin{figure}
\centering
\includegraphics[width=10cm,height=8cm]{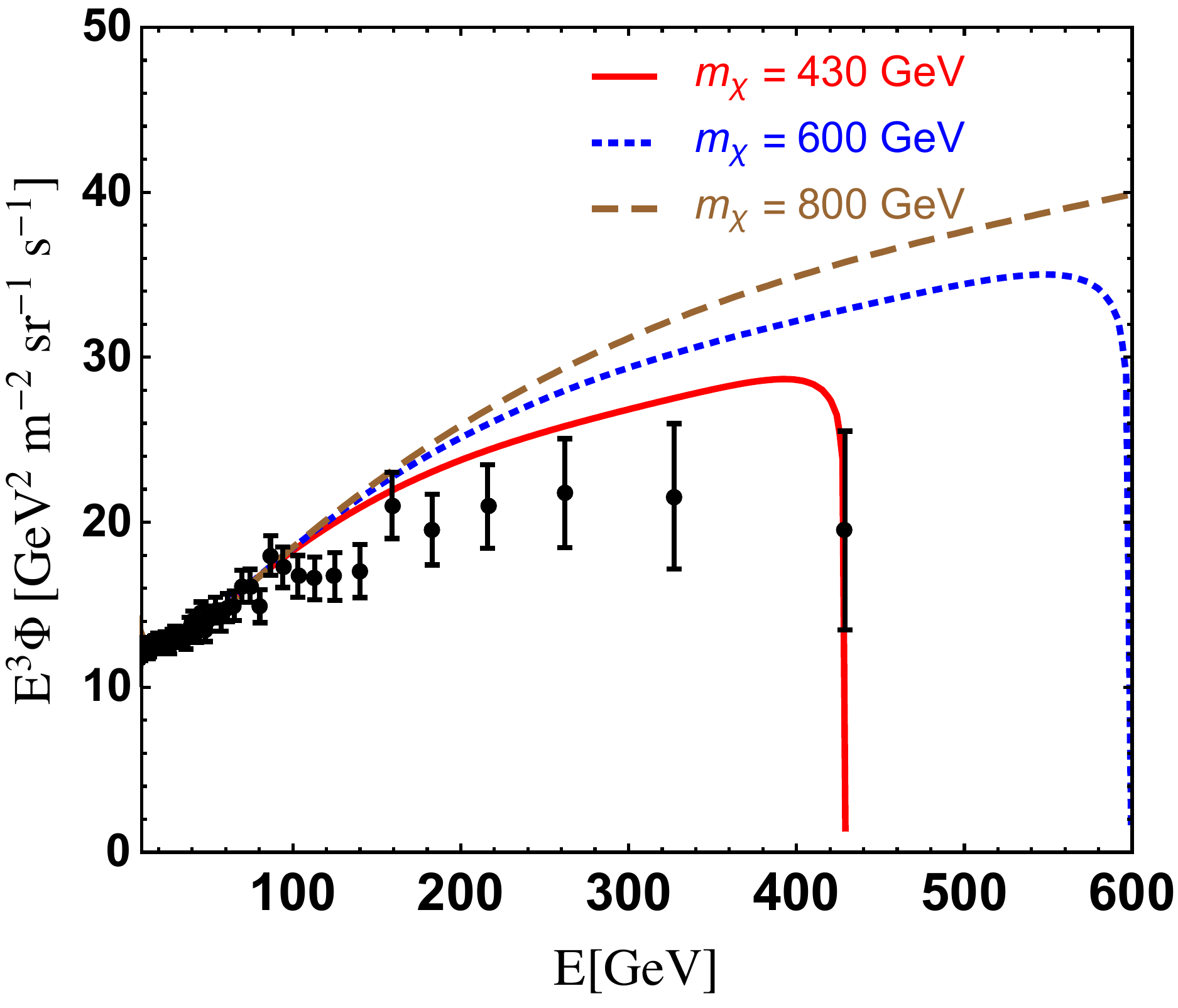}
\caption{\label{fig:positron flux} The predictions of positron fluxes from the $\bar \chi \chi \to Z' \to \ell^+ \ell^-$ processes for different DM masses and the $M_{Z'}=1\,\TeV$ case together with the AMS-02 measurements of the positron fluxes~\cite{Aguilar:2014mma}.}
\end{figure}

With the primary and secondary positron fluxes given in Eqs.~(\ref{eq:prime_positron_flux}) and (\ref{eq:second_positron_flux}) respectively, we combine them into the total positron flux as
\beqn\label{eq:tot_positron_flux}
\Phi_+^{\rm tot}(E)&=&\Phi_+^{\rm prime}(E)\times {\rm BF}+ \Phi_+^{\rm sec}(E)\,,
\eeqn
where a boost factor (BF) is included. This BF parametrizes the effects of local DM clumps distributions in the area where DM annihilation takes place. As shown in Eq.~\eqref{eq:positron_source}, enhancing the local density by a factor of $10$ would yield an enhancement of the positron flux by a factor of $100$. Afterwards, we perform the following $\chi^2$ fit 
\beqn\label{eq:chi2}
\chi^2&=&\sum_i \frac{(\mu_i^{\rm model} - \mu_i^{\rm AMS})^2}{\sigma_i^2}\,,
\eeqn
to the AMS-02 data of positron fluxes in order to determine the corresponding BF for each benchmark model in Table.~\ref{tab:DM_benchmark}. The summation in Eq.~\eqref{eq:chi2} goes through each energy bin for the positron spectrum from $10\,\GeV$ up to $430\,\GeV$. Below $\sim 10 \,\GeV$, the counting rate of the positrons becomes unstable, which is affected by the solar modulation~\cite{Aguilar:2014mma}. Here $\mu_i^{\rm model}$ are the products of $E^3\cdot \Phi_+^{\rm tot}(E)$ with the total positron fluxes given in Eq. (\ref{eq:tot_positron_flux}). $\mu_i^{\rm AMS}$ and $\sigma_i$ are the AMS-02 measurement of the positron fluxes $E^3\cdot \Phi$ and errors in the energy range of $E\in (10\,,430)\,\GeV$. As an example, for the $m_\chi=430\,\GeV$ and $M_{Z'}=1\,\TeV$, we get the minimum of $\chi^2/{\rm d.o.f}\simeq 0.87$ when ${\rm BF}=330$\,. We demonstrate the positron flux predictions from the benchmark model with $m_\chi = 430\,\GeV$ and $M_{Z'}=1\,\TeV$ in Fig.~\ref{fig:positron flux}. The models with different values of $M_{Z'}$ predict the same positron flux once some appropriate BF's are taken into account. To fit the AMS-02 predictions, we expect a large boost factor (${\rm BF}\sim 100$) for large mass of DM. The fits to other benchmark models and the corresponding BF's are listed in Table.~\ref{tab:DM_benchmark} previously.

Now, we give some comments on our fit to the AMS-02 positron flux. First, our fit is based on the assumption that the $Z'$ couples with the three generations of leptons equally. In general, the positron spectrums generated from the $e, \mu, \tau$ final states have different shapes. If the fraction of each lepton in the final state is not fixed, but determined after fitting to the data, then there would be a much better agreement between the experimental data and the theoretical prediction from DM annihilation~\cite{Lin:2014vja,Cao:2014cda, Jin:2014ica}. Second, we notice that the present experimental result only cover the energy spectrum of the positron flux up to about 500 GeV. As a result, the DM of a mass of about 430 GeV would be able to fit this spectrum. At the same time, because of the mass limit, the positron spectrum induced by the DM is ended at about 430 GeV. In the future, it is likely that the AMS-02 data show a continuing existence of the positron flux over a larger energy range. In order to account for this possibility, we also show the fit with DM masses of 600 and 800 GeV in Fig.~\ref{fig:positron flux}. It is seen that they start to deviate from the present data, which means that the DM of a mass heavier than $600\,\GeV$ with equal couplings to leptons is less likely to explain the current AMS-02 data in this leptophilic $Z'$ model.


\section{\hspace*{-2mm}The Collider Searches for $Z'$ and Its Invisible Decays}
\label{section:collider_search}

The leptophilic DM with $Z'$ interaction is just one of the many models trying to explain the positron fluxes observed. This explanation can be tested by other experiments, such as the collider searches. Given that the vector boson $Z'$ is not too heavy, it is likely to produce $Z'$ directly at the LHC and the next generation $e^+ e^-$ colliders. The search for leptophilic $Z'$ at the hadron colliders is challenging due to the large QCD background, where the jets can be misidentified as leptons. In this work, we focus on the search at the next generation $e^+ e^-$ colliders. We expect that not only a heavy $Z'$ is discovered, but also its coupling with the DM can be confirmed.

\subsection{The decays of $Z'$ and its searches at the $e^+ e^-$ colliders}

\begin{figure}
\centering
\includegraphics[width=8cm,height=8cm]{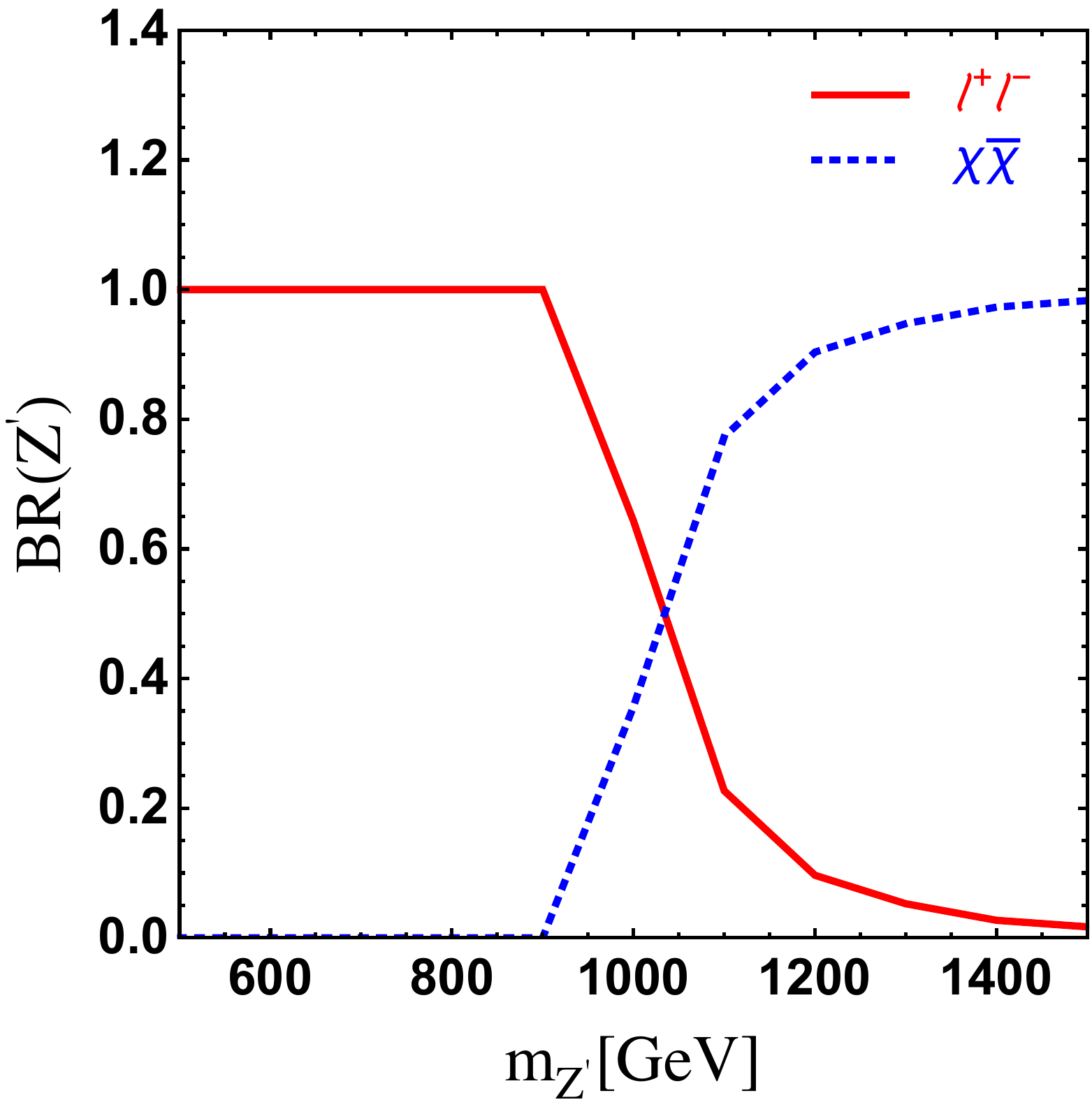}
\caption{\label{fig:BRZp} The ${\rm BR}[Z' \to \ell^+ \ell^-]$ (red curve) and ${\rm BR}[Z' \to \bar \chi \chi]$ (blue curve) for $M_{Z'}\in (0.5\,\TeV\,, 1.5\,\TeV)$.}
\end{figure}

The vector boson $Z'$ couples to both SM leptons and Dirac DM $\chi$. The partial decay width to the SM leptons is
\beqn\label{eq:Zpvis_wid}
\Gamma[Z' \to \ell^+ \ell^- ]&=& \frac{  M_{Z'} }{12\pi} \sqrt{ 1- \frac{4 m_\ell^2 }{M_{Z'}^2 } } \Big[ (g_\ell^V)^2 (1+ 2\frac{m_\ell^2}{M_{Z'}^2}) + (g_\ell^A)^2  (1 - \frac{4m_\ell^2}{M_{Z'}^2 })  \Big]\,.
\eeqn
Under the massless limit of $m_\ell=0$ and the simplified couplings of $\lambda_\ell=g_\ell^V = g_\ell^A$, one has this visible partial decay widths become
\beqn
\Gamma[Z' \to \ell^+ \ell^- ]&=& \frac{ M_{Z'}}{6\pi }\lambda_\ell^2 \,.
\eeqn
The invisible decay width to the Dirac DM $\chi$ reads:
\beqn\label{eq:Zpinvis_wid}
\Gamma[Z'\to \bar \chi \chi]&=& \frac{ M_{Z'} }{12\pi} \sqrt{ 1- \frac{4 m_\chi^2 }{M_{Z'}^2 } } \Big[ (g_\chi^V)^2 (1+ 2\frac{m_\chi^2}{M_{Z'}^2}) + (g_\chi^A)^2  (1 - \frac{4m_\chi^2}{M_{Z'}^2 })  \Big]\non
&=&\frac{M_{Z'}}{6\pi} \lambda_\chi^2 \sqrt{ 1- \frac{4 m_\chi^2 }{M_{Z'}^2 } } (1 - \frac{m_\chi^2}{M_{Z'}^2})\,,
\eeqn
with the coupling simplifications of $\lambda_\chi = g_\chi^V = g_\chi^A$ in the second line. In Fig.~\ref{fig:BRZp}, we show the decay branching fractions ${\rm BR}[Z']$ by combining the visible and invisible decay widths in Eqs.~(\ref{eq:Zpvis_wid}) and (\ref{eq:Zpinvis_wid}). The evaluations are made for the benchmark models given in Table.~\ref{tab:DM_benchmark}. For the mass range of $M_{Z'}\in (0.5\,\TeV\,, 1.0\,\TeV)$, the decay mode of $Z'\to \ell^+ \ell^-$ dominates. When the invisible decay channel of $Z' \to \bar \chi \chi$ is kinematically allowed, i.e., $M_{Z'}\gtrsim 2m_\chi$, the invisible decay mode becomes more dominant.

Below, we will study the leptophilic $Z'$ searches at the future high-energy $e^+ e^-$ colliders based on the benchmark models that we listed in the previous Table.~\ref{tab:DM_benchmark}. Considering the mass ranges and interaction properties of $Z'$, we are interested in the $e^+ e^-$ colliders with center-of-mass energy $\sqrt{s}\gtrsim 1\,\TeV$. Therefore, we explore the experimental searches at the ILC $1\,\TeV$ run and the CLIC $1.5\,\TeV$ run. In practice, we implement the benchmark models into UFO model files by using {\tt FeynRules}~\cite{Christensen:2008py} for the event generation. Events are generated for both signals and the corresponding SM background processes by {\tt Madgraph}~\cite{Alwall:2014hca}. Afterwards, the {\tt PYTHIA}~\cite{Sjostrand:2006za} and {\tt PGS}~\cite{PGS} will be used for the parton shower, hadronization, and fast detector simulations. For the detector simulations, we implement the design of the ILD~\cite{Abe:2010aa} in the {\tt PGS}. For both ILC and CLIC runs, we consider the polarization scheme of $P(e^+\,,e^-)=(-30\,\% \,, +80\,\%)$ for the incoming electron and positron beams, which will suppress the SM background contributions efficiently~\cite{Yu:2013aca,Yu:2014ula}

\subsection{The direct searches for the $Z'$ through the di-lepton plus photon channel}

We first discuss the searches for $Z'$ through the OSSF di-lepton associated with a single photon channel: $e^+ e^- \to Z' (\to \ell^+ \ell^-) \gamma $, where $\ell=(e\,,\mu)$. In particular, we require that the $Z'$ is produced on-shell. For the benchmark models we listed in Table.~\ref{tab:DM_benchmark}, we will study the signals for the $M_{Z'}\in (0.5\,\TeV\,, 0.8\,\TeV)$ samples at the ILC $1\,\TeV$ run, and the $M_{Z'}\in (1.0\,\TeV\,, 1.5\,\TeV)$ samples at the CLIC $1.5\,\TeV$ runs. The relevant SM background processes contributing to the OSSF di-lepton plus photon final states include: $e^+ e^- \to \ell^+ \ell^- \gamma$, $e^+ e^- \to W^+ W^- \gamma $, and $e^+ e^- \to \tau^+ \tau^- \gamma$. It turns out only the first process of $e^+ e^- \to \ell^+ \ell^-\gamma$ contribute dominantly, while the cross sections for all other processes are sub-dominant. In addition, events containing jets and large $\MET$ from the $e^+ e^- \to W^+ W^- \gamma $ and $e^+ e^- \to \tau^+ \tau^- \gamma$ processes can be vetoed, hence the contributions from these two processes can be safely neglected.

\begin{figure}
\centering
\includegraphics[width=7cm,height=5cm]{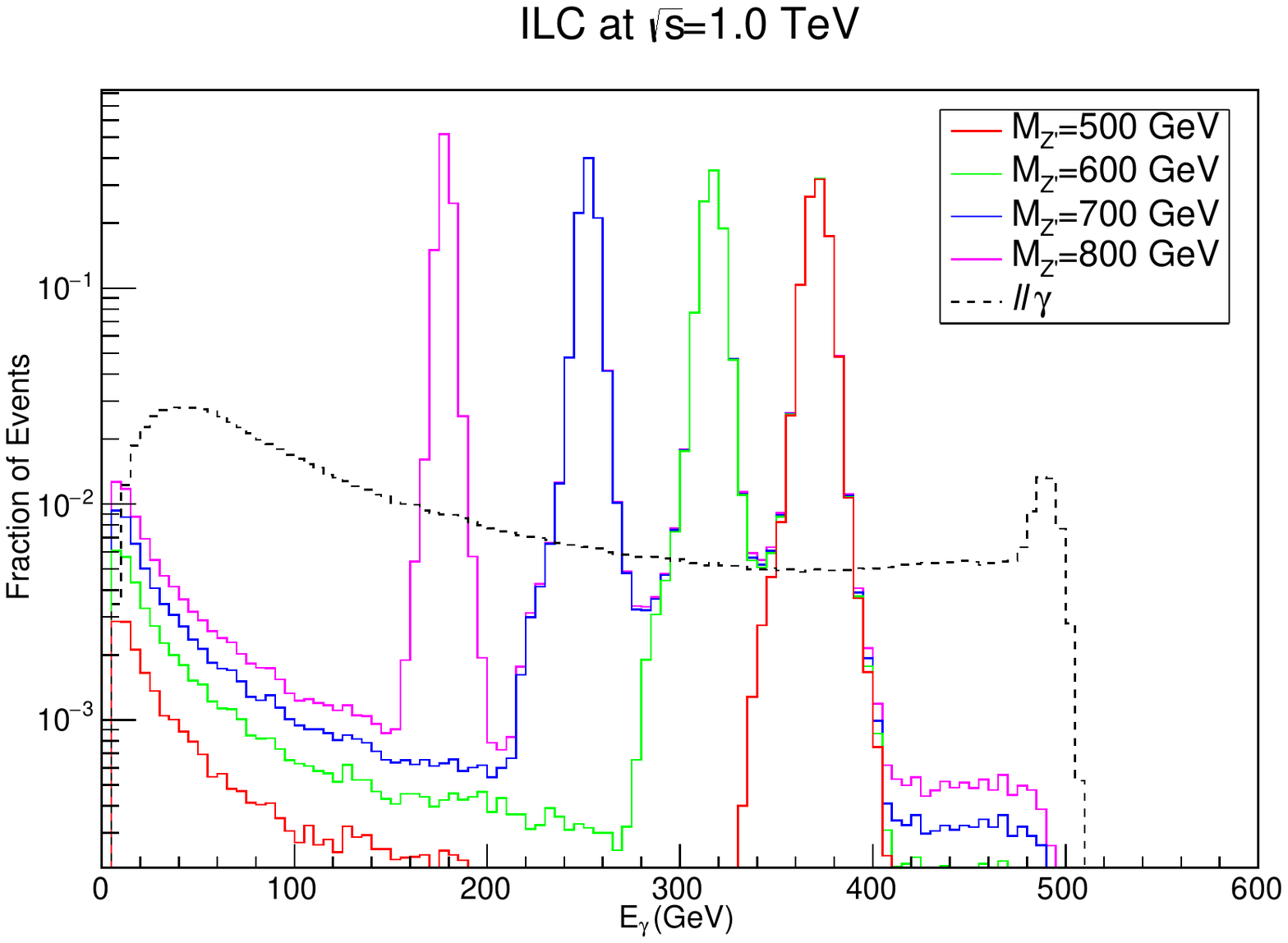}
\includegraphics[width=7cm,height=5cm]{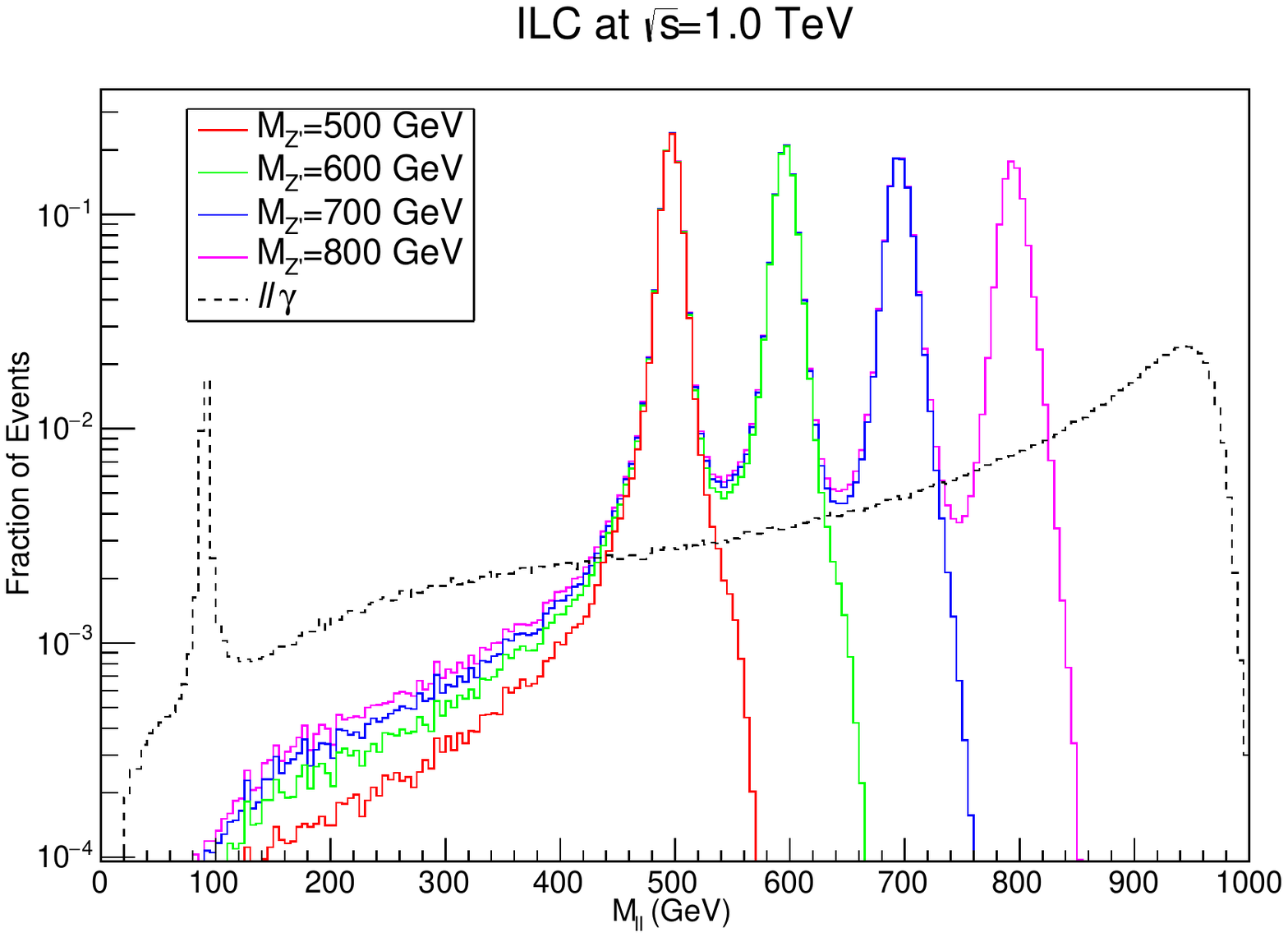}
\caption{\label{fig:Zpllgamma} The distributions of the single photon energy $E_\gamma$ (left panel) and the OSSF di-lepton invariant masses (right panel) for the signal process of $e^+ e^- \to Z'(\to \ell^+ \ell^-) \gamma$ and the SM background process of $e^+ e^- \to \ell^+ \ell^- \gamma$ at the ILC $1\,\TeV$ runs.}
\end{figure}

For illustration, we list a set of kinematic cuts to be imposed at the ILC $1\,\TeV$ runs for the benchmark models with $M_{Z'}\in (0.5\,\TeV\,, 0.8 \,\TeV)$.
\begin{itemize}

\item Cut-1: all events containing jets with $p_T\geq 10\,\GeV$ and $|\eta|\leq 2.5$ are vetoed. Besides, we also veto events with $\MET \geq 50\,\GeV$.

\item Cut-2: the events containing OSSF di-lepton plus a photon are selected. The OSSF di-leptons should satisfy the requirements: $|\eta(\ell^\pm)|\leq 2.5$, $p_T(\ell_0)\geq 20\,\GeV$, $p_T(\ell_1)\geq 10\,\GeV$, and $2.0\leq \Delta R(\ell^+ \ell^-)\leq 5.0$. Here $\ell_0$ and $\ell_1$ represent the leading and sub-leading charged leptons ordered by their transverse momenta.

\item Cut-3: The photon in the selected events satisfy the requirements: $|\eta(\gamma)|\leq 2.5$ and $p_T(\gamma)\geq 20\,\GeV$. Since the photon is produced via the $e^+ e^- \to \gamma Z'$ process, the photon energy is determined by $(s - M_{Z'}^2)/2\sqrt{s}$. Considering the detector smearing effects, we impose the following cuts to the single photon energy $E_\gamma$:
\beqn\label{eq:ILC_Egamma_cut}
&& \frac{s - M_{Z'}^2}{2\sqrt{s}} - 25\,\GeV \leq E_\gamma \leq \frac{s - M_{Z'}^2}{2\sqrt{s}} + 25\,\GeV\,.
\eeqn
The corresponding $E_\gamma$ distributions for the signal and the SM background processes are shown in the left-panel of Fig.~\ref{fig:Zpllgamma}.

\item Cut-4: finally we reconstruct the invariant mass of the OSSF di-leptons around the $Z'$ mass window: $|M_{\ell\ell} - M_{Z'} |\leq 50\,\GeV$. The distributions of the invariant mass of OSSF di-leptons $M_{\ell\ell}$ for the signal and the SM background processes are shown in the right-panel of Fig.~\ref{fig:Zpllgamma}.

\end{itemize}
To search for the benchmark models with $M_{Z'}\in (1.0\,\TeV\,, 1.5\,\TeV)$ at the CLIC $1.5\,\TeV$ runs, we shall follow the similar cut flows listed above for the ILC $1\,\TeV$ runs. Meanwhile, we modify the single photon energy cut window from $25\,\GeV$ in Eq.~\eqref{eq:ILC_Egamma_cut} to $50\,\GeV$. Furthermore, the cut to the invariant mass window of OSSF di-leptons $M_{\ell\ell}$ will be modified to $|M_{\ell\ell} - M_{Z'} |\leq 100\,\GeV$ for the CLIC $1.5\,\TeV$ runs.

\begin{table}[htb]
\begin{center}
\begin{tabular}{c|c|c|c|c|c|c}
\hline
  & cut-0 & cut-1 & cut-2 & cut-3 & cut-4  & $5\,\sigma$    \\\hline\hline
 $M_{Z'}=0.5\,\TeV$ & $138$ & $124$ & $118$ & $114$ & $111$ & $145\,\pb^{-1}$  \\\hline
 $\ell^+ \ell^- \gamma$ & $1,125$ & $1,050$ & $619$ & $39$ & $31$ & $-$  \\\hline
 \hline
  $M_{Z'}=0.6\,\TeV$ & $215$ & $191$ & $187$ & $180$ & $175$ & $65\,\pb^{-1}$  \\\hline
   $\ell^+ \ell^- \gamma$ & $1,125$ & $1,050$ & $619$ & $40$ & $35$ & $-$  \\\hline
 \hline
 $M_{Z'}=0.7\,\TeV$ & $238$ & $210$ & $205$ & $198$ & $192$ & $61\,\pb^{-1}$  \\\hline
  $\ell^+ \ell^- \gamma$ & $1,125$ & $1,050$ & $619$ & $46$ & $41$ & $-$  \\\hline
 \hline
  $M_{Z'}=0.8\,\TeV$ & $447$ & $395$ & $384$ & $373$ & $360$ & $27\,\pb^{-1}$  \\\hline
 $\ell^+ \ell^- \gamma$ & $1,125$ & $1,050$ & $619$ & $59$ & $54$ & $-$  \\\hline
 \hline
\end{tabular}
\caption{\label{tab:Zpllgamma_cuts_ILC} The cross sections (unit: $\fb$) after imposing kinematic cuts for the benchmark models from the OSSF $\ell^+\ell^-$ plus single photon channel and the relevant SM background processes at the ILC $1.0\,\TeV$ runs. In the last column for different benchmark model signals, we list the integrated luminosity needed for the $5\,\sigma$ excess via the OSSF di-lepton channel. }
\end{center}
\end{table}

\begin{table}[htb]
\begin{center}
\begin{tabular}{c|c|c|c|c|c|c}
\hline
& cut-0 & cut-1 & cut-2 & cut-3 & cut-4  & $5\,\sigma$    \\\hline\hline
 $M_{Z'}=1.0\,\TeV$ & 265 & 228 & 222 & 216 & 211 & $43\,\pb^{-1}$ \\\hline
 $\ell^+ \ell^- \gamma$ & 594 & 553 & 339 & 29 & 27 & $-$  \\\hline
 \hline
 $M_{Z'}=1.1\,\TeV$ & 119 & 101 & 99 & 93 & 91 & $159\,\pb^{-1}$ \\\hline
 $\ell^+ \ell^- \gamma$ & 594 & 553 & 313 & 30 & 28 & $-$  \\\hline
 \hline
 $M_{Z'}=1.2\,\TeV$ & 78 & 65 & 63 & 57 & 55 & $409\,\pb^{-1}$  \\\hline
 $\ell^+ \ell^- \gamma$ & 594 & 553 & 313 & 35 & 33 & $-$  \\\hline
  \hline
 $M_{Z'}=1.3\,\TeV$ & 69 & 57 & 55 & 46 & 44  & $762\,\pb^{-1}$  \\\hline
 $\ell^+ \ell^- \gamma$ & 594 & 553 & 313 & 48 & 45 &  $-$  \\\hline
  \hline
 $M_{Z'}=1.4\,\TeV$ & 66 & 54 & 51 & 37 & 35 & $1.55\,\fb^{-1}$  \\\hline
 $\ell^+ \ell^- \gamma$  & 594 & 553 & 313 & 70 & 64 & $-$  \\\hline
  \hline
 $M_{Z'}=1.5\,\TeV$ & 25 & 20 & 19 & 4 & 4 & $22.5\,\fb^{-1}$  \\\hline
 $\ell^+ \ell^- \gamma$ & 594 & 553 & 313 & 15 & 13 & $-$  \\\hline
\end{tabular}
\caption{\label{tab:Zpllgamma_cuts_CLIC} The cross sections (unit: $\fb$) after imposing kinematic cuts for the benchmark models from the OSSF $\ell^+\ell^-$ plus single photon channel and the relevant SM background processes at the CLIC $1.5\,\TeV$ runs. In the last column for different benchmark model signals, we list the integrated luminosity needed for the $5\,\sigma$ excess via the OSSF di-lepton channel. }
\end{center}
\end{table}

After imposing the series of kinematic cuts, we list the cut efficiencies for the benchmark models at the ILC $1\,\TeV$ runs and at the CLIC $1.5\,\TeV$ runs in Table.~\ref{tab:Zpllgamma_cuts_ILC} and Table.~\ref{tab:Zpllgamma_cuts_CLIC} respectively. The integrated luminosities required to observe $5\,\sigma$ excess from the OSSF di-lepton plus a single photon channel for each benchmark model are also listed. For benchmark models of $M_{Z'}\in (0.5\,\TeV\,, 0.8\,\TeV)$, the $Z'$ can be searched via the $e^+ e^- \to Z'(\to \ell^+ \ell^-) \gamma$ channel when the $1\,\TeV$ ILC runs with the integrated luminosities reaching $\mO(10)\,\pb^{-1}$. For benchmark models of $M_{Z'}\in (1.0\,\TeV\,, 1.5\,\TeV)$, the signal channel of $e^+ e^- \to Z'(\to \ell^+ \ell^-) \gamma$ can reach $5\sigma$ C.L. at the $1.5\,\TeV$ CLIC runs with the integrated luminosities reaching $\mO(10)\,\pb^{-1} - \mO(10)\,\fb^{-1}$.

\subsection{The direct searches for the invisible modes of $Z'$: mono-photon channel}

Besides the direct discovery of the leptophilic $Z'$ at the future $e^+ e^-$ colliders, here we discuss the search for the invisible decay modes of $Z'$ via the mono-photon process of $e^+ e^- \to Z'(\to \bar \chi \chi) \gamma$. The previous studies on the DM searches at the $e^+ e^-$ colliders and the implications to the DM direct and indirect searches were carried out in Refs.~\cite{Birkedal:2004xn, Borodatchenkova:2005ct, Fox:2011fx, Dreiner:2012xm,Yu:2013aca}. For the mono-photon channel, we should look for the $\gamma+\MET$ signal. The corresponding SM background should be: $e^+ e^- \to \bar \nu \nu \gamma$, which are due to the $e^+ e^- \to Z(\to \bar\nu \nu) \gamma$ and the $t-$channel $W^\pm-$exchanging processes. Other processes like $e^+ e^- \to \ell^+ \ell^- \gamma$ and $e^+ e^- \to \tau^+ \tau^- \gamma$ are also likely to fake the $\gamma+\MET$ signal, when neither of the charged leptons or jets are tagged in the detector. The cross sections for the relevant SM background processes at the CLIC $1.5\,\TeV$ runs with the beam polarizations of $P(e^+ \,, e^-)=(-30\%\,, +80\%)$ read
\beqs\label{eqs:monophoton_SMbkg}
\beqn
&&\sigma[e^+ e^- \to \bar \nu \nu \gamma]=  408\,\fb \,,\\
&&\sigma[e^+ e^- \to \ell^+ \ell^- \gamma]= 594\,\fb \,,\\
&& \sigma [e^+ e^- \to \tau^+ \tau^- \gamma] = 13\,\fb \,.
\eeqn
\eeqs
Again, the cross sections are evaluated by considering the preliminary cuts imposed at the parton level by {\tt Madgraph}.

\begin{figure}
\centering
\includegraphics[width=7.5cm,height=5cm]{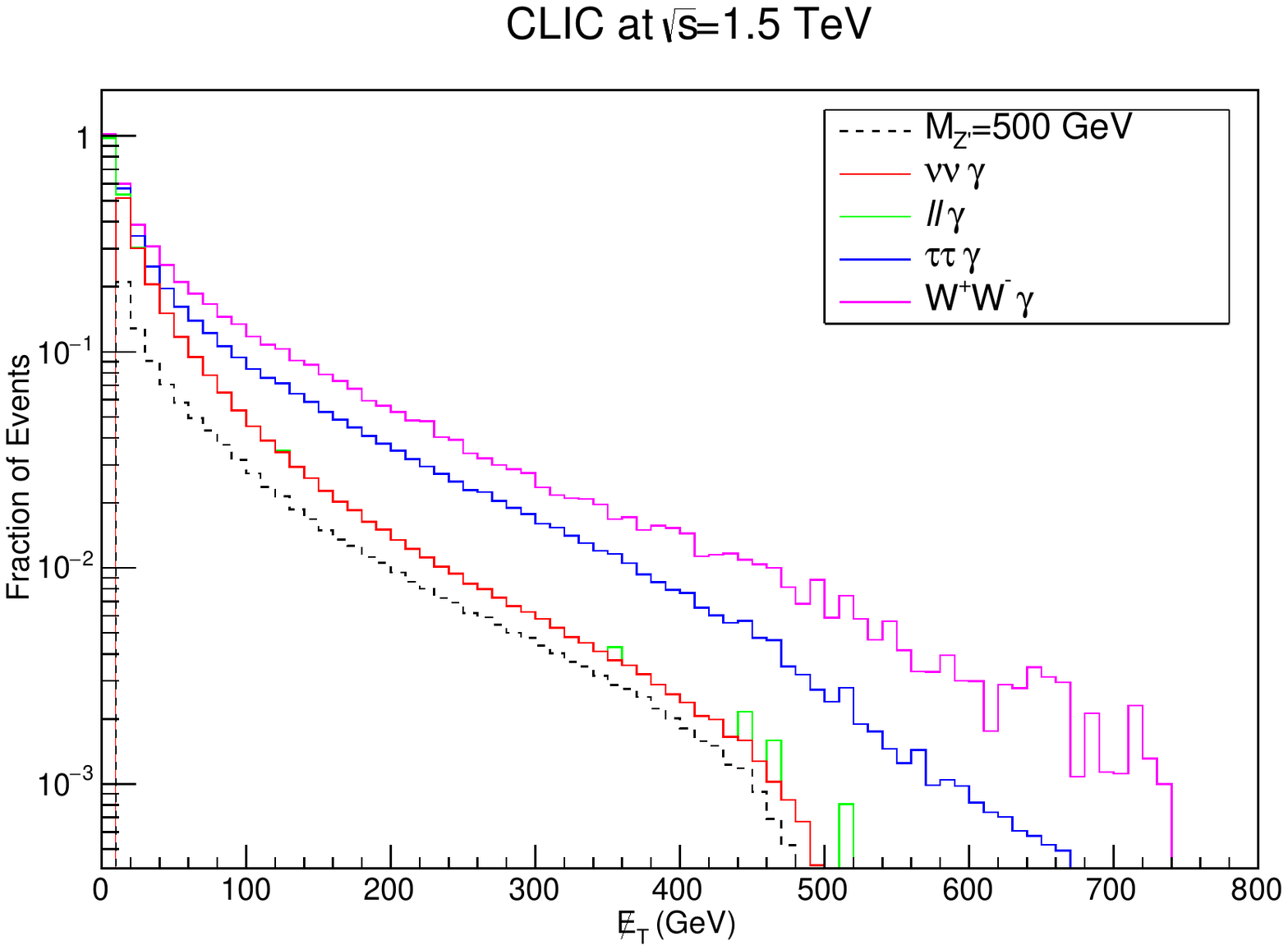}
\includegraphics[width=7.5cm,height=5cm]{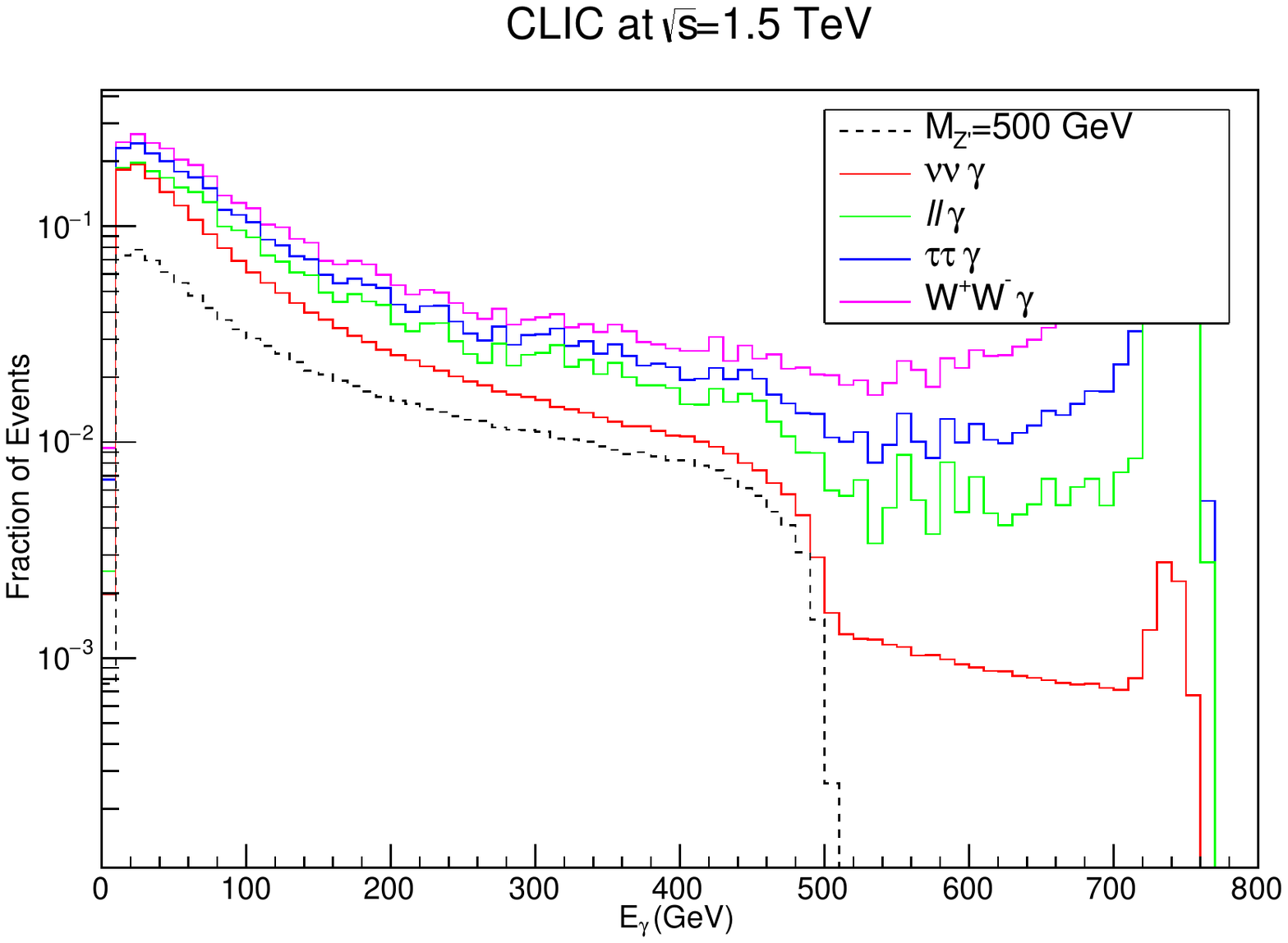}
\caption{\label{fig:Zp500_METEgamma} The distributions of the $\MET$ (left-panel) and the single photon energy $E_\gamma$ (right-panel) in the mono-photon signal process of $e^+ e^- \to Z' (\to \bar \chi \chi) \gamma$ and the corresponding SM background processes. The benchmark model with $M_{Z'}=500\,\GeV$ is taken for the CLIC $1.5\,\TeV$ runs.}
\end{figure}

\begin{table}[htb]
\begin{center}
\begin{tabular}{c|c|c|c|c|c}
\hline
  & cut-0 & cut-1 & cut-2  & cut-3 & $5\sigma$   \\\hline\hline
 $M_{Z'}=0.5\,\TeV$ & $9.7$ & $9.4$ & $9.3$ & $9.3$ & $113\,\fb^{-1}$   \\\hline
  $M_{Z'}=0.6\,\TeV$ & $6.1$ & $5.9$ & $5.8$ & $5.8$ & $290\,\fb^{-1}$  \\\hline
 $M_{Z'}=0.7\,\TeV$ & $4.3$ & $4.2$ & $4.1$ & $4.1$ & $580\,\fb^{-1}$ \\\hline
 $M_{Z'}=0.8\,\TeV$ & $0.99$ & $0.98$ & $0.94$ & $0.94$ & $11\,\ab^{-1}$ \\\hline
 $\bar \nu \nu \gamma$ & $408$ & $404$ & $401$ & $390$ & $-$   \\\hline
 \hline
\end{tabular}
\caption{\label{tab:monoA_cuts00} The cross sections (unit: $\fb$) after each kinematic cut to be imposed for the benchmark models in the mono-photon process and the relevant SM background processes at the CLIC $1.5\,\TeV$ runs. }
\end{center}
\end{table}

Now we discuss imposing the appropriate kinematic cuts in order to select signal events from the SM background processes. For the benchmark models with $500\,\GeV \leq M_{Z'}\leq 800\,\GeV$, the mono-photon process of $e^+ e^- \to \bar \chi \chi \gamma$ is produced through the off-shell exchanges of the $Z'$. Accordingly, we illustrate the set of kinematic cuts to be imposed:
\begin{itemize}

\item Cut-1: the events containing jets and leptons with $p_T\geq 10\,\GeV$ and $|\eta|\leq 2.5$ are vetoed. The events containing $\gamma+\MET$ signal will be selected, where the single photon should further satisfy the conditions: $p_T(\gamma)\geq 10\,\GeV$ and $|\eta(\gamma)|\leq 2.5$.

\item Cut-2: the events with $\MET\leq 400\,\GeV$ are selected, and the corresponding distributions of the $\MET$ for both signals and SM background processes are shown in the left panel of Fig.~\ref{fig:Zp500_METEgamma} after imposing the Cut-1.

\item Cut-3: the single photon energies in this set of benchmark models should be bounded from above: $E_\gamma \leq (s - 4 m_\chi^2 )/(2\sqrt{s})\simeq 503\,\GeV$ for the CLIC $1.5\,\TeV$ runs. The corresponding distributions of the single photon energy $E_\gamma$ for both signals and SM background processes are displayed in the right panel of Fig.~\ref{fig:Zp500_METEgamma} after imposing the Cut-1 and Cut-2. We also note an equivalent kinematic variable to consider is the recoil mass of:
\beqn\label{eq:rec_mass}
m_{\rm rec}&\equiv& \sqrt{ (p_{e^+} + p_{e^-} - p_\gamma)^2 }\,.
\eeqn
For the set of benchmark models with $M_{Z'}\in (0.5\,\TeV\,,0.8\,\TeV)$, one expects that $m_{\rm rec}\geq 2m_\chi =860\,\GeV$. The cut of the recoil mass turns out to be equivalent to the $E_\gamma$ cut here.

\end{itemize}
Afterwards, we list the kinematic cut efficiencies in Table.~\ref{tab:monoA_cuts00} after imposing Cut-1 to Cut-3 for the benchmark models with $M_{Z'}\in (0.5\,\TeV\,,0.8\,\TeV)$. The searches of the mono-photon channel for this set of benchmark models require the integrated luminosities of $\sim\mO(100)\,\fb^{-1}$ or even larger, which are generally challenging. This is due to the smallness of the mono-photon cross section with an off-shell invisible decay mode $Z'\to \bar\chi \chi$ produced.

\begin{table}[htb]
\begin{center}
\begin{tabular}{c|c|c|c|c|c|c}
\hline
  & cut-0 & cut-1 & cut-2  & cut-3 & cut-4 & $5\sigma$ \\\hline\hline
 $M_{Z'}=1.0\,\TeV$ & $240.6$ & $240.0$ & $239.9$ & $233.5$ & $231.6$ & $21\,\pb^{-1}$ \\\hline
 $\bar \nu \nu \gamma$ & $408$ & $404$ & $402$ & $9.3$ & $6.1$ & $-$  \\\hline
 \hline
 $M_{Z'}=1.1\,\TeV$ & $672$ & $665$ & $664$ & $619$ & $608$ & $7\,\pb^{-1}$ \\\hline
 $\bar \nu \nu \gamma$ & $408$ & $404$ & $401$ & $14.1$ & $10.3$ & $-$ \\\hline
 \hline
 $M_{Z'}=1.2\,\TeV$ & $1,173$ & $1,153$ & $1,148$ & $1,022$ & $1,001$ & $4\,\pb^{-1}$ \\\hline
 $\bar \nu \nu \gamma$ & $408$ & $404$ & $397$ & $23.8$ & $19.0$ & $-$  \\\hline
 \hline
 $M_{Z'}=1.3\,\TeV$ & $2,010$ & $1,963$ & $1,926$ & $1,643$ & $1,608$ & $3\,\pb^{-1}$ \\\hline
 $\bar \nu \nu \gamma$ & $408$ & $404$ & $387$ & $48.8$ & $42.2$ & $-$ \\\hline
 \hline
 $M_{Z'}=1.4\,\TeV$ & $3,819$ & $3,703$ & $3,681$ & $3,202$ & $3,147$ & $2\,\pb^{-1}$ \\\hline
 $\bar \nu \nu \gamma$ & $408$ & $404$ & $375$ & $161$ & $150$ & $-$ \\\hline
 \hline
  $M_{Z'}=1.5\,\TeV$ & $2,264$ & $2,108$ & $2,084$ & $1,759$ & $1,747$ & $4\,\pb^{-1}$ \\\hline
 $\bar \nu \nu \gamma$ & $408$ & $404$ & $348$ & $161$ & $158$ & $-$ \\\hline
 \hline
\end{tabular}
\caption{\label{tab:monoA_cuts} The cross sections (unit: $\fb$) after each kinematic cut to be imposed for the benchmark models of $M_{Z'}\in (1.0\,\TeV\,, 1.5\,\TeV)$ in the mono-photon process and the relevant SM background processes at the CLIC $1.5\,\TeV$ runs. }
\end{center}
\end{table}

For the benchmark models with $M_{Z'}\in (1.0\,\TeV \,, 1.5\,\TeV)$, the invisible decays of $Z'\to \bar \chi \chi$ occurs on-shell. We impose the following set of cuts to distinguish the signals from the SM background:
\begin{itemize}

\item Cut-1: the events with $\gamma+\MET$ are selected. The selected photon should satisfy the conditions:  $|\eta(\gamma)| \leq 2.5$, $p_T(\gamma)\geq 10\,\GeV$. The events with leptons or jets of $p_T(\ell\,,j)\geq 10\,\GeV$ and $|\eta(\ell\,,j)|\leq 2.5$ are vetoed. 

\item Cut-2: we set an upper bound to the $\MET$ of the selected events:
\beqs
\beqn
M_{Z'}=1.0\,\TeV&:&  0\,\GeV\leq  \MET\leq 450\,\GeV\,,\\
M_{Z'}=1.1\,\TeV&:& 0\,\GeV \leq \MET \leq 400\,\GeV \,,\\
M_{Z'}=1.2\,\TeV&:& 0\,\GeV \leq \MET \leq 300\,\GeV \,,\\
M_{Z'}=1.3\,\TeV&:& 0\,\GeV \leq \MET \leq 200\,\GeV \,,\\
M_{Z'}=1.4\,\TeV&:& 0\,\GeV \leq \MET \leq 150\,\GeV \,,\\
M_{Z'}=1.5\,\TeV&:& 0\,\GeV \leq \MET \leq 100\,\GeV \,.
\eeqn
\eeqs

\item Cut-3: the single photon energy in the signal events is still determined by $E_\gamma=(s-M_{Z'}^2)/2\sqrt{s}$. As for the CLIC $1.5\,\TeV$ case, we determine the single photon energy as: $E_\gamma^{\rm CLIC} \in (E_\gamma - 25\,\GeV\,, E_\gamma + 25\,\GeV)$.

\item Cut-4:  the recoil mass $m_{\rm rec}$ defined in (\ref{eq:rec_mass}) can be used to reconstruct the $Z'$ mass. Thus, we require the cut of $|m_{\rm rec} - M_{Z'}| \leq 50 \,\GeV$ for the $M_{Z'}\in (1\,\TeV\,, 1.5\,\TeV)$ samples.

\end{itemize}
In Table.~\ref{tab:monoA_cuts}, we list the cross sections for the $\gamma+\MET$ signals and the dominant SM background of $\bar \nu \nu \gamma$ after each kinematic cut imposed. For the CLIC running at $\sqrt{s}=1.5\,\TeV$, our analysis shows that the mono-photon signal searches for the benchmark models can reach $5\,\sigma$ excess with the integrated luminosities of $\sim\mO(1)\,\pb^{-1}$.


\vspace*{3mm}
\section{\hspace*{-2mm}Conclusions}
\label{section:conclusion}

The current indirect DM search results are intriguing for probing the DM natures and their possible connection to the visible sector described by the SM. The search for the cosmic positrons from both PAMELA and the recent AMS-02 collaborations keep showing the disagreement between the measurements and the known astrophysical predictions. The DM candidates annihilating into various SM final states were considered as possible source for such excessive high-energy positron fluxes. In this work, we hypothesize the couplings between Dirac DM particle $\chi$ and the charged leptons $\ell_i$ in the SM mediated by a leptophilic $Z'$. After considering the experimental constraints from muon anomalous magnetic moment, and the LEP-II searches, we show that this setup is capable of fitting the current AMS-02 results of the positron fluxes.

Such a setup with a leptophilic $Z'$ can be clearly tested in the future high-energy $e^+ e^-$ collider, with the center-of-mass energy running at $\mO(1)\,\TeV$. The searches for the $Z'$ is considered for the OSSF di-lepton plus a single photon final state. We analyze the signals for the benchmark models and the corresponding SM background contributions, and find the discovery of the $Z'$ requires moderate luminosities at ILC/CLIC runs. Furthermore, the searches for the DM productions are essential for determining the DM couplings to $Z'$. For the benchmark models with $Z'$ mass of $M_{Z'}\in (1.0\,\TeV\,, 1.5\,\TeV)$, the searches for the invisible decays of $Z'\to \bar\chi\chi$ is easily achieved at the CLIC $1.5\,\TeV$ runs via the mono-photon process. Meanwhile, for the light mediator cases of $M_{Z'}\in (0.5\,\TeV\,, 0.8\,\TeV)$, the mono-photon cross sections are typically small, making the discovery of the invisible mode challenging.


\section*{Acknowledgments}

We would like to thank Qing-Hong Cao, Ti Gong, Jia Liu, Alexander Pukhov, Qi-Shu Yan, and Zhao-Huan Yu for useful discussions and communications. N.C. is partially supported by National Science Foundation of China (under grant No. 11335007). The research of J.W. has been supported by the Cluster of Excellence {\it Precision Physics, Fundamental Interactions and Structure of Matter} (PRISMA-EXC 1098).


\end{document}